\documentclass[12pt,twoside,a4paper]{article}
\usepackage{amsmath,amssymb,latexsym,theorem,bbm,natbib,color,url,booktabs}
\usepackage[FIGTOPCAP]{subfigure}
\usepackage{multirow,graphicx}  
\newfont{\msa}{msam10 scaled\magstep1}
\newfont{\ssmsa}{msam9}

\def\crps{\mathop{\hbox{\rm CRPS}}}

\numberwithin{equation}{section}

\title{Similarity-based semi-local estimation of EMOS models}

\author{{\sc Sebastian Lerch$^{1}$} and {\sc S\'andor Baran$^{2}$}\\[1.1em]
         $^1$Heidelberg Institute for Theoretical Studies\\
         Schloss-Wolfsbrunnenweg 35, D-69118 Heidelberg, Germany\\
         $^2$Faculty of Informatics, University of Debrecen\\
         Kassai \'ut 26, H-4028 Debrecen, Hungary 
         }

\date{\today}

\begin{document}
\pagestyle{myheadings}

\maketitle

\begin{abstract}

Weather forecasts are typically given in the form of forecast ensembles obtained from multiple runs of numerical weather prediction models with varying initial conditions and physics parameterizations. Such ensemble predictions tend to be biased and underdispersive and thus require statistical postprocessing. In the ensemble model output statistics (EMOS) approach, a probabilistic forecast is given by a single parametric distribution with parameters depending on the ensemble members. 
This article proposes two semi-local methods for estimating the EMOS coefficients where the training data for a specific observation station are augmented with corresponding forecast cases from stations with similar characteristics. Similarities between stations are determined using either distance functions or clustering based on various features of the climatology, forecast errors, ensemble predictions and locations of the observation stations.
In a case study on wind speed over Europe with forecasts from the Grand Limited Area Model Ensemble Prediction System, the proposed similarity-based semi-local models show significant improvement in predictive performance compared to standard regional and local estimation methods. They further allow for estimating complex models without numerical stability issues and are computationally more efficient than local parameter estimation.

\end{abstract}

\section{Introduction}
  \label{sec:sec1}
  
Many applications such as agriculture, wind energy production or aviation require accurate and reliable forecasts of wind speed. Wind speed predictions are usually based on output from numerical weather prediction (NWP) models which describe the dynamical and physical behavior of the atmosphere through nonlinear partial differential equations.

Historically, single runs of NWP models with the best available initial conditions were used to obtain single-valued predictions of the future state of the atmosphere. However, such deterministic forecasts fail to account for uncertainties in the initial conditions and the numerical model. Therefore, NWP models are nowadays often run several times with varying initial conditions and/or numerical representations of the atmospheric processes, resulting in an ensemble of forecasts \citep{gr05, lp}. Since the first operational implementation by the European Centre for Medium-Range Weather Forecasts \citep[ECMWF, see][for a description of the current version]{ecmwf}, the generation of ensemble forecasts has become standard practice in meteorology. All major national meteorological services operate their own ensemble prediction systems (EPSs) as for example the PEARP\footnote{PEARP: Pr\'evision d'Ensemble ARPege} EPS of M\'eteo France \citep{dljbac} or the COSMO-DE\footnote{COSMO: Consortium for Small-scale 
Modeling} EPS of the German Meteorological Service \citep{btg}.

Recent developments in ensemble forecasting include multi-model ensemble prediction systems such as the THORPEX Interactive Grand Global Ensemble \citep[TIGGE,][]{tigge15} where several single-model ensembles each based on multiple runs of individual NWP models are combined, see, e.g., \citet{js09,hbhlp12}. Another example is the Grand Limited Area Model Ensemble Prediction System \citep[GLAMEPS,][]{iversen11} considered in this article which is described in more detail in Section \ref{sec:sec2}. 

Generally, probabilistic forecasts, i.e., forecasts given in the form of full probability distributions, are desirable as they allow for a quantification of the uncertainty associated with the prediction. Probabilistic forecasts further allow for optimal decision making since optimal deterministic forecasts can be obtained as functionals of the forecast distributions \citep{gneiting11}. This is particularly important for applications such as wind power forecasting for auction processes in electricity markets where the optimal bidding strategy depends on permanently changing features of the market conditions \citep{pck07,pinson13}.

While the implementation of ensemble prediction systems is an important step in the transition from deterministic to probabilistic forecasting, ensemble forecasts are finite and do not provide full predictive distributions. Further, ensemble forecasts generally tend to be underdispersive and subject to systematic bias, and thus require some form of statistical postprocessing \citep{hc97,gr05}.

Various methods for statistical postprocessing of ensemble forecasts have been developed over the last years, for recent reviews and comparisons, see, e.g., \citet{sk10,rs12,wfk,gneiting14}. State of the art techniques include Bayesian model averaging \citep[BMA;][]{rgbp} and ensemble model output statistics (EMOS) or non-homogeneous regression \citep{grwg}. Both approaches provide estimates of the future distributions of the weather variables of interest and are partially implemented in the {\tt ensembleBMA} and {\tt ensembleMOS} packages for the statistical programming language {\tt R} \citep{frgsb}.

In the case of BMA, the predictive probability density function (PDF) of a future weather quantity is a weighted mixture of individual PDFs corresponding to the members of the ensemble, where the weights are determined by the relative performance of the ensemble members during a given training period. The BMA models for various weather quantities differ in the PDFs of the mixture components. For wind speed, \citet{sgr10} suggest the use of a gamma mixture, whereas \citet{bar} considers BMA component PDFs following a truncated normal (TN) rule. 

The EMOS approach is conceptually simpler, the predictive PDF is given by a single parametric distribution with parameters depending on the ensemble members. Over the last years, EMOS models have been developed for  calibrating ensemble forecasts of various weather variables such as temperature and sea level pressure \citep{grwg}, wind speed \citep{tg,lt,bl} and precipitation \citep{sch}.

The parameters of the forecast distributions are typically estimated by minimizing proper scoring rules evaluated at forecasts and verifying observations over rolling training periods consisting of the preceding $n$ days \citep{gneiting14}. For selecting the corresponding training sets, two basic approaches  are given by local and regional methods. In the local approach, only forecast cases from the single observation station of interest are considered for the parameter estimation, whereas in the regional approach, data from all available observation stations are composited to form a single training set for all stations. Local estimation generally results in better predictive performance \citep[see, e.g.,][]{tg,stg}, however, is often problematic if only limited amounts of training data are available. On the other hand, there are typically no numerical stability issues in regional parameter estimation, however, in case of large ensemble domains it is undesirable to obtain a single 
set of coefficients for all observation stations due to the potentially significant differences in the climatological properties of the observation stations and forecast errors of the ensemble.

We apply the truncated normal EMOS model of \citet{tg} for statistical postprocessing of wind speed forecasts of the 52-member GLAMEPS ensemble. The GLAMEPS ensemble covers a large domain across Europe and Northern Africa, however, only a short period of data is available. 

We propose two similarity-based semi-local approaches to parameter estimation in order to account for these challenges. A distance-based approach uses data from stations with similar characteristics to augment the training data for a given stations and follows ideas of \citet{hhw}. Our novel clustering-based approach employs $k$-means clustering to obtain groups of similar observation stations with respect to various features which then form shared training sets for parameter estimation. 

The remainder of this article is organized as follows. In Section \ref{sec:sec2}, we introduce the GLAMEPS ensemble and the observation data. In Section \ref{sec:sec3}, we review the truncated normal EMOS model and propose similarity-based semi-local approaches to parameter estimation based on distance functions and clustering. In Section \ref{sec:sec4}, we report the results of the case study based on the GLAMEPS data. We conclude with a discussion in Section \ref{sec:sec5}.

\section{The GLAMEPS ensemble}
  \label{sec:sec2}

\begin{figure}[t]
\centering
\subfigure[$\hspace{-1.25cm}$]{\includegraphics[width=0.49\textwidth]{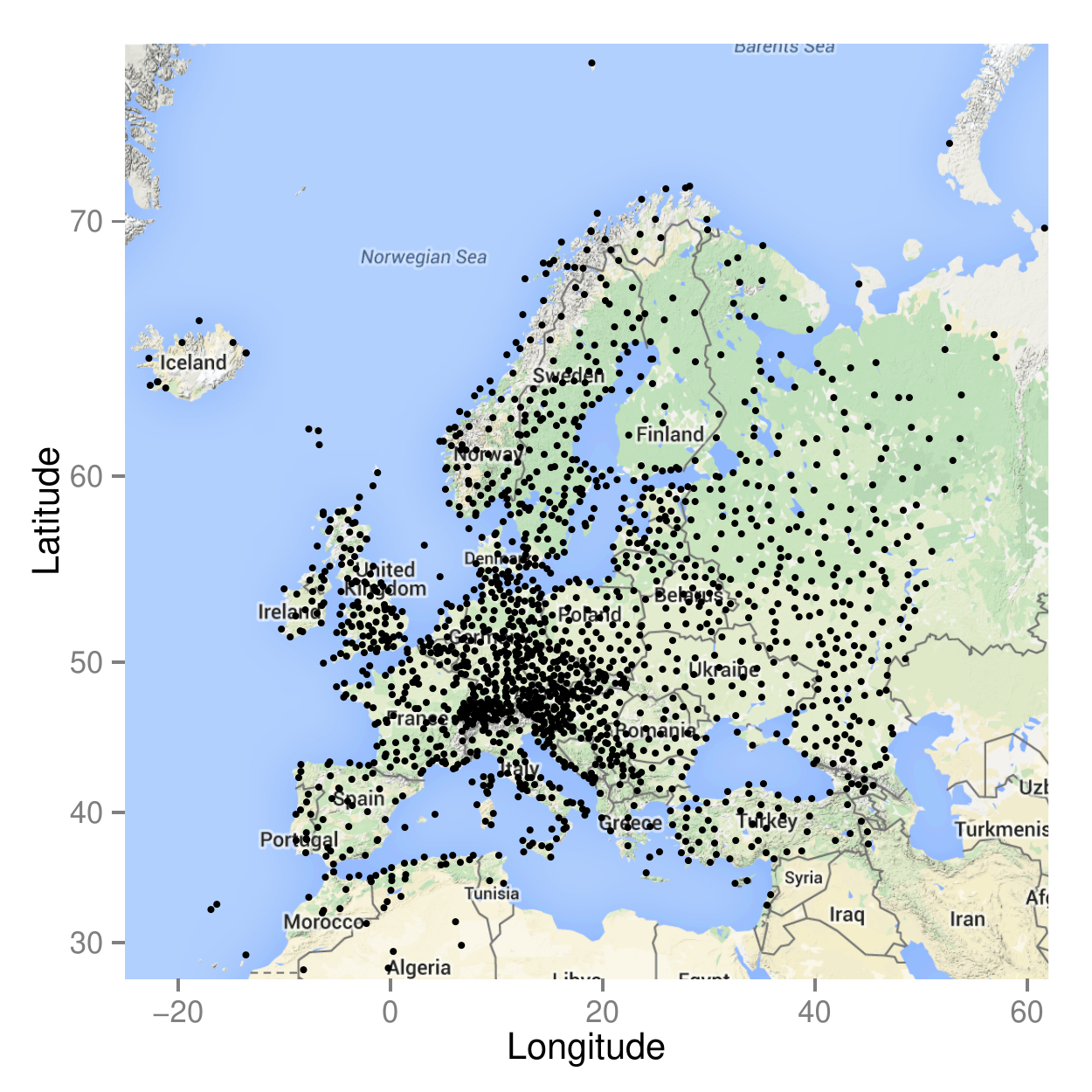}}\hfill  
\subfigure[$\hspace{-1.25cm}$]{\includegraphics[width=0.49\textwidth]{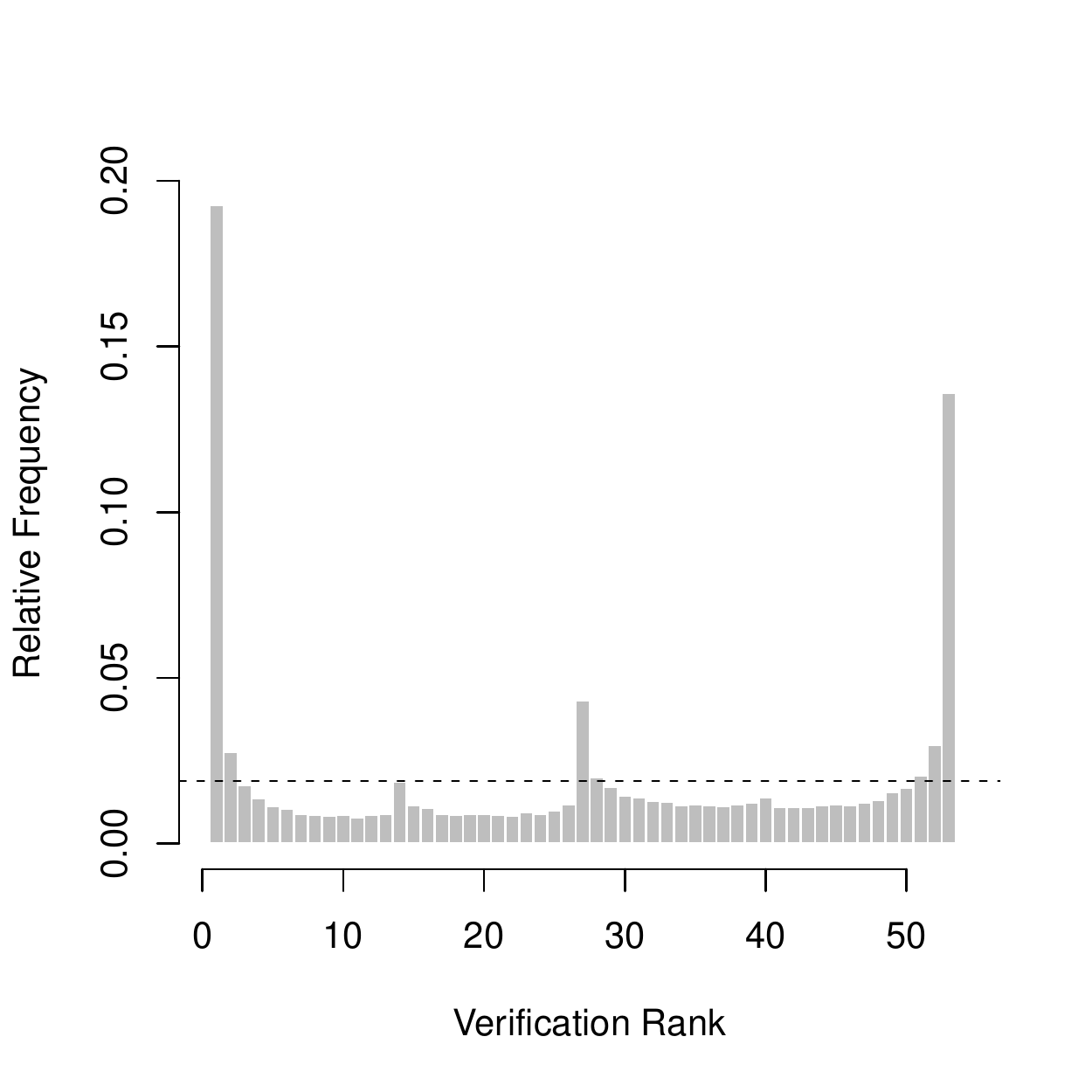}}\hfill  
\caption{Locations of observation stations (a) and verification rank histogram (b) of the GLAMEPS ensemble. The map in (a) and all following maps in this article were produced using the {\tt ggmap} package for {\tt R} \citep{kw}.}
\label{fig:fig1}
\end{figure}

The GLAMEPS ensemble is a short-range multi-model EPS launched in 2006 as a part of the cooperation between the ALADIN\footnote{Aire Limit\'ee Adaptation dynamique Developpement International} and HIRLAM\footnote{High Resolution Limited Area Modelling} consortia. It operates on a large domain covering Europe, North-Africa and the Northern Atlantic and the currently running Version 2 (GLAMEPSv2) is a combination of the subensembles from two versions of the ALARO model (ISBA and SURFEX schemes, see, e.g., \citet{np} and \citet{surfex}) and two version of the HIRLAM model (Kain-Fritsch and STRACO schemes, see, e.g., \citet{kf} and \citet{sass}). Each subensemble consists of $12$ perturbed members and a control forecast, and half of the perturbed members are lagged by 6h \citep{deckmyn}.

Our data base contains 52 ensemble members of 18h ahead forecasts of 10-m wind speed for 1738 observation sites (see Figure \ref{fig:fig1}a) together with the corresponding validating observations for October 2 -- November 25, 2013, and February 2 -- May 18, 2014. We divide the available data into two equally large periods from October 2013 to February 2014 and from March 2014 to May 2014 in order to allow for rolling training periods of sufficient length. The forecasts are evaluated over the second period. Data from the first period are used to obtain training periods of equal lengths for all days, and to estimate the similarities between the stations used in the distance-based semi-local approach to parameter estimation, see Section \ref{sec:sec3} for details. While \citet{iversen11} apply BMA to calibrate temperature forecasts of the GLAMEPS ensemble, the article at hand is the first application of postprocessing techniques to the corresponding wind speed forecasts to the best of the authors' knowledge.

Figure \ref{fig:fig1}b shows the verification rank histogram of the raw ensemble. This is the histogram of ranks of validating observations relative to the corresponding 52 ensemble member forecasts over the verification period \citep[see, e.g.,][Section 7.7.2]{wilks}. For a calibrated ensemble, forecasts and observations should be exchangeable and all observed ranks should thus be equally likely and follow a uniform distribution corresponding to the dashed line in Figure \ref{fig:fig1}b. The U-shaped verification rank histogram of the GLAMEPS ensemble indicates that the GLAMEPS forecasts lack calibration and are underdispersive, i.e., too many observations fall outside the ensemble range. This deficiency can be observed for various ensemble prediction systems, see, e.g., \citet{bl}.

\section{Ensemble model output statistics}
  \label{sec:sec3}

As discussed in the Introduction, the goal of ensemble postprocessing is to correct for biases and dispersion errors in NWP model output. The EMOS approach uses a single parametric distribution to model the PDF of the future weather quantity, where the parameters depend on the ensemble members. In case of the wind speed this PDF should be concentrated on the non-negative values. Here we apply the truncated normal EMOS model introduced by \citet{tg}, however, alternative EMOS models utilizing a generalized extreme value distribution \citep{lt} and a log-normal distribution \citep{bl} are available and have also been tested. As these alternative choices do not offer substantial improvements for the data at hand, we limit our discussion to results for the TN model and note that similar conclusions apply for the alternative EMOS approaches. 

\subsection{Truncated normal EMOS models}
  \label{subs:subs3.1}

The PDF of the TN distribution with location $\mu$, scale $\sigma>0$, and cut-off at zero, denoted by ${\mathcal N}_{\,0}\big(\mu,\sigma^2\big)$, is given by
\begin{equation*}
g(x\vert\, \mu,\sigma):=\frac{\frac
  1{\sigma}\varphi\big((x-\mu)/\sigma\big)}{\Phi\big(\mu/\sigma\big)
}, \quad x\geq 0, \qquad \text{and} \qquad g(x\vert\, \mu,\sigma):=0,
\quad \text{otherwise,}
\end{equation*}
where $\varphi$ and $\Phi$ are the PDF and the cumulative distribution function (CDF) of the standard normal distribution, respectively. The EMOS predictive distribution proposed by \citet{tg} is 
\begin{equation}
   \label{eq:eq3.2}
  {\mathcal N}_0\big(a_0+a_1f_1+ \cdots +a_Mf_M,b_0+b_1S^2\big) \qquad
  \text{with} \qquad S^2:=\frac 1{M-1}\sum_{k=1}^M\big (f_k-\overline f\big)^2,
\end{equation}
where $f_1,f_2,\dots ,f_M$ denote the ensemble of distinguishable forecasts of wind speed for a given location and time, and $\overline f$ denotes the ensemble mean. Location parameters $a_0, a_1, \dots, a_M$ and scale parameters $b_0,b_1$  of model \eqref{eq:eq3.2} can be estimated from the training data consisting of ensemble forecasts and verifying observations from the preceding $n$ days by optimizing an appropriate verification score (see Section \ref{subs:subs3.2}). 

However, in case of the GLAMEPS ensemble, similar to the majority of the currently used ensemble prediction systems such as the ECMWF ensemble or the PEARP EPS of M\'eteo France, some of the ensemble members are generated with the help of perturbations of the initial conditions simulating model uncertainties. This should be incorporated into the model formulation since these exchangeable members are assumed to be statistically indistinguishable.

In what follows, if we have $M$ ensemble members divided into $m$ groups of exchangeable members, where the $k$th group contains $M_k\geq 1$ ensemble members ($\sum_{k=1}^mM_k=M$), notation $f_{k,\ell}$ is used for the  $\ell$th member of the $k$th group. In this situation  ensemble members within a given group share the same coefficient of the location parameter \citep{frg, gneiting14} resulting in the predictive distribution
\begin{equation}
   \label{eq:eq3.3}
  {\mathcal N}_0\bigg(a_0+a_1\sum_{\ell_1=1}^{M_1}f_{1,\ell_1}+ \cdots
  +a_m\sum_{\ell_m=1}^{M_m} f_{m,\ell_m},b_0+b_1S^2\bigg), 
\end{equation}
where again, $S^2$ denotes the ensemble variance. Model formulations that take into account the grouping in modeling the variance have also been investigated, but result in a reduction of the predictive performance \citep{bl}.

\subsection{Verification scores}
  \label{subs:subs3.2}
  
 As argued concisely by \citet{gbr}, the general goal of probabilistic forecasting should be to maximize the sharpness of the predictive distribution subject to calibration. While calibration is a notion of statistical consistency between the predictive distribution and the observation, sharpness is a property of the forecasts only and refers to the information content in the forecast distribution. 
Calibration of EMOS post-processed forecasts can be assessed using  probability integral transform (PIT) histograms. The PIT is the value of the predictive CDF evaluated at the verifying observations \citep{rgbp} and  the closer the histogram to the desired uniform distribution, the better the calibration. PIT histograms can be seen as continuous analogues of verification rank histograms, see Section \ref{sec:sec2}.

Further, one can also investigate the coverage of the central prediction interval corresponding to the nominal coverage of the raw ensemble which is $51/53$ or $96.2\,\%$ for the GLAMEPS ensemble. 
The coverage of a $(1-\alpha)100 \,\%, \ \alpha \in (0,1),$ central prediction interval is the proportion of validating observations located between the lower and upper $\alpha/2$ quantiles of the predictive
distribution. For a calibrated probabilistic forecast this value should be around $(1-\alpha)100 \,\%$ and the choice of $\alpha$ corresponding to the nominal coverage allows direct comparison to the raw ensemble. Given the predictive distribution is calibrated, it should be as sharp as possible, where sharper distributions correspond to narrower central prediction intervals.

Proper scoring rules assign numerical values to pairs of forecasts and observations and can be used to assess calibration and sharpness simultaneously \citep{grjasa}. The most popular scoring rules providing summary measures of predictive performance are the logarithmic score, i.e., the negative logarithm of the predictive PDF evaluated at the verifying observation, and the continuous ranked probability score \citep[CRPS;][]{grjasa,wilks}. Given a predictive CDF $F(y)$ and an observation $x$, the CRPS is defined as 
\begin{equation}
\crps\big(F,x\big):=\int_{-\infty}^{\infty}\big (F(y)-{\mathbbm 
  1}{\{y \geq x\}}\big )^2{\mathrm d}y={\mathsf E}|X-x|-\frac 12
{\mathsf E}|X-X'|, 
\end{equation}
where ${\mathbbm 1}\{H\}$ denotes the indicator of a set $H$, while $X$ and $X'$ are independent random variables with CDF $F$ and finite first moment. In case of ensemble forecasts, the predictive CDF is given by the empirical CDF of the ensemble. The CRPS can be expressed in the same unit as the observation and both scores are proper scoring rules which are negatively oriented, i.e. smaller scores indicate better predictive performance.  

Following the optimum score estimation approach of \citet{grjasa}, proper scoring rules can be utilized in parameter estimation by minimizing the average value of a proper scoring rule over a training set. In this way the optimization with respect to the logarithmic score corresponds to the classical maximum likelihood (ML) estimation of the parameters. In case of a truncated normal predictive distribution the CRPS has a closed form \citep[see, e.g.,][]{tg} which allows for an efficient parameter estimation based on optimizing the mean CRPS. 

Point forecasts given by the median value of the predictive distribution are evaluated using the mean absolute error (MAE) quantifying the deviation from the corresponding validating observations to assess the deterministic predictive accuracy. Note that the median value is the optimal point forecast under the MAE \citep{gneiting11,pinhag}.

\subsection{Similarity-based semi-local parameter estimation}
  \label{subs:subs3.3}

In general, the coefficients of the TN EMOS model are estimated by minimizing the mean CRPS of the predictive distributions over suitably chosen rolling training periods consisting of the preceding $n$ days. There exist two basic approaches for selecting the training data \citep{tg,stg}. The regional (or global) approach composites ensemble forecasts and validating observations from all available stations during the rolling training period. Therefore, one obtains a single universal set of parameters across the entire ensemble domain, which is then used to produce the forecasts at all observation sites. In case of the GLAMEPS ensemble this means that a single set of coefficients is used for the wide-ranging domain and the geographical and climatological variability might thus not be sufficiently taken into account. While the regional approach to parameter estimation can be implemented without numerical stability issues and offers slight gains in predictive performance compared to the raw ensemble (see Section 
\ref{sec:sec4}), there is room for further improvement for large and heterogeneous domains.

By contrast, the local approach produces distinct parameter estimates for different stations by using only the training data of the given station. Local models typically result in better predictive performance compared to regional models \citep[see, e.g.,][]{tg,stg}, however, these training sets contain only one observation per day and the estimation of local EMOS models thus requires significantly longer training periods to avoid numerical stability issues. For example, in model \eqref{eq:eq3.3} with $12$ exchangeable groups (which is the case for the GLAMEPS ensemble, see Section \ref{sec:sec4}) the number of free parameters to be estimated is $15$, making the use of local EMOS impossible for small data sets such as the one considered in this article. In a recent case study on EMOS models for the ECMWF ensemble, \citet{hemri14} find that training period lengths between 365 and 1816 days give the best results for local parameter estimation. For the GLAMEPS data at hand, choosing such long training periods 
is impossible as the whole data set consists of only 161 days.

We propose two alternative similarity-based semi-local approaches which avoid the problems that make both regional and local estimation of the EMOS coefficients undesirable for the GLAMEPS data. The basic idea of the semi-local methods is to combine the advantages of regional and local estimation by augmenting the training data for a given station with data from stations with similar characteristics. The choice of similar stations is either based on suitably defined distance functions, or on clustering.

\subsubsection*{Distance-based semi-local model}

Following \citet{hhw}, the training sets of a given station are increased by including training data from other stations with similar features. The similarity between stations is determined based on suitably defined distance functions\footnote{We use the term \emph{distance function} in a general sense with only one of the proposed similarity measures depending on the actual geographical locations of the observation stations. From a mathematical point of view, all considered distance functions are semimetrics, i.e. non-negative and symmetric functions $d:\mathbb{R}\times\mathbb{R} \rightarrow \mathbb{R}$ with $d(i,i) = 0$. Distance functions can thus be seen as negatively oriented similarity measures with smaller values indicating more similar characteristics of the stations of interest.}. Note that compared to \citet{hhw}, we consider alternative choices of distance functions, and our forecasts are evaluated over a set of observation stations whereas the forecasts and analysis data used by \citet{hhw} are 
given on a grid. Different conclusions may apply for grid-based data.

Generally, the distance between two stations $i$ and $j$ denoted by $d(i,j)$ with $i,j\in\{1,\dots,1738\}$ is determined using the first period of available data from October 2013 to February 2014 which is distinct from the verification period. In the semi-local estimation of the EMOS model for a given station $i_0$, we then add the corresponding forecast cases in the rolling training period from the $L$ most similar stations, i.e., the $L$ stations with the smallest distances $d(i_0,j), \ j\in\{1,\dots,1738\}$. 

Alternatively, one could also iteratively determine the similarities anew in every rolling training period. However, this approach requires lots of computational resources as all pair-wise distances between stations have to be re-computed for every training period (up to symmetry), and is thus infeasible due to the large number of observation stations. In particular, note that already the simple distance-based semi-local model estimation with a fixed set of distances is computationally more demanding compared to local parameter estimation which arises as special case for $L = 1$. Furthermore, initial tests did not indicate significant improvements in the predictive performance for the GLAMEPS data, we thus limit our discussion to the use of the first period of data for determining the similarities between stations for the distance-based approach. 

We investigate the following five distance functions.

{\em Distance 1: Geographical locations}. The distance between stations $i$ and $j$ is given by the Euclidean distance of the locations $(\mathcal{X}_i,\mathcal{Y}_i)$ and $(\mathcal{X}_j,\mathcal{Y}_j)$ of the two stations, i.e., 
\[
 d^{(1)}(i,j) := \sqrt{(\mathcal{X}_i - \mathcal{X}_j)^2 + (\mathcal{Y}_i - \mathcal{Y}_j)^2}.
\]
The Euclidean distance is employed here since the station locations in the data set are given on the linearly transformed model estimation grid. In general, the spherical or great-circle distance is a more appropriate distance measure for actual geographical locations on the globe. 

{\em Distance 2: Station climatology}. Let $\hat F_i$ denote the empirical CDF of wind speed observations at station $i$ over the first period of data. Similar to the distance function proposed by \citet{hhw}, the distance to station $j$ is given by the normalized sum over the absolute differences of the respective empirical CDFs $\hat F_i$ and $\hat F_j$ evaluated at a set of fixed values $S$, i.e., 
\[
 d^{(2)}(i,j) := \frac{1}{|S|} \sum_{x\in S} \left| \hat F_i(x) - \hat F_j(x) \right|,
\]
where $|S|$ denotes the cardinality of $S$. 
Here, we choose $S = \{0, 0.5, 1, 1.5, \dots, 14.5, 15\}$ and note that the obtained similarities are robust to minor changes in the definition of $S$. 

{\em Distance 3: Ensemble forecast errors}. Denote the ensemble mean for station $i$ and date $t,$ by $\bar f_{i,t}$ and the corresponding verifying observation by $x_{i,t},$ then the forecast error $e_{i,t}$ of the ensemble mean is given by 
\[
 e_{i,t} = \bar f_{i,t} - x_{i,t}.
\]
The third distance function is based on the distribution of these forecast errors. To that end, we define the empirical CDF of the forecast errors at station $i$ as 
\begin{equation}\label{eq:ECDFfcerrors}
 \hat G_i^e(z) := \frac{1}{|T|} \sum_{t \in T} \mathbbm{1} \{ \bar f_{i,t} - x_{i,t} \leq z \},
\end{equation}
where $T $ denotes the set of dates in the first period of data. The distance between two stations $i$ and $j$ is then given by
\[
 d^{(3)}(i,j) :=  \frac{1}{|S^\prime|} \sum_{x\in S^\prime} \left| \hat G_i^e(x) - \hat G_j^e(x) \right|,
\]
where $S^\prime = \{ -10, -9.5, -9, -8.5, \dots, 0 , \dots, 8.5, 9, 9.5, 10\}$ denotes the set of fixed values at which the empirical CDFs of the forecast errors are evaluated. As before, the obtained sets of similar stations are robust to changes of $S^\prime$.

{\em Distance 4: Combination of distance 2 and 3}. We add up the values of distances 2 and 3 to define a distance function which depends on both the climatology of the observations as well as the distribution of the forecast errors of the ensemble, i.e., with the above notation,
\begin{equation*}
 d^{(4)}(i,j) := d^{(2)}(i,j) + d^{(3)}(i,j) =  \frac{1}{|S|} \sum_{x\in S} \left| \hat F_i(x) - \hat F_j(x) \right| +   \frac{1}{|S^\prime|}\sum_{x\in \tilde{S}^\prime} \left| \hat G_i^e(x) - \hat G_j^e(x) \right|.
\end{equation*}

{\em Distance 5: Ensemble characteristics}. \citet{schefzik15} proposes a similarity-based implementation of the Shaake shuffle using a distance function that depends on summary statistics of the ensemble. With $\bar f_{i,t}$ and  $S_{i,t}$ denoting the mean and standard deviation of the ensemble member forecasts at station $i$ and date $t$, the distance between station $i$ and $j$ is given by
\[
 d^{(5)}(i,j) := \sum_{t\in T} \sqrt{ \left(\bar f_{i,t} - \bar f_{j,t} \right)^2 + \left( S_{i,t} - S_{j,t} \right)^2 },
\]
where $T$ again denotes the set of dates during the first period of data.

\begin{figure}[t]
\centering
\subfigure[$\hspace{-1.25cm}$]{\includegraphics[width=0.49\textwidth]{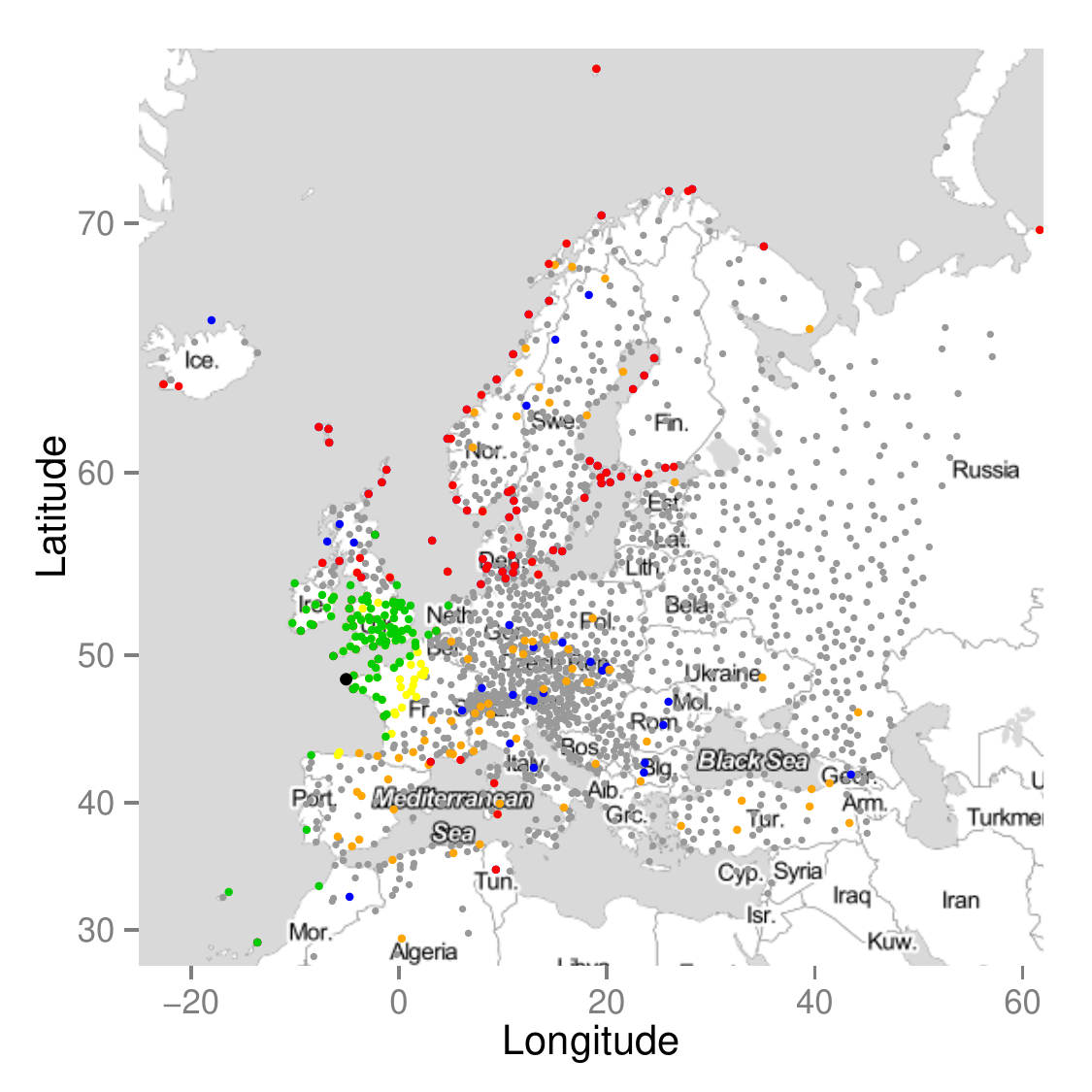}}\hfill  
\subfigure[$\hspace{-1.25cm}$]{\includegraphics[width=0.49\textwidth]{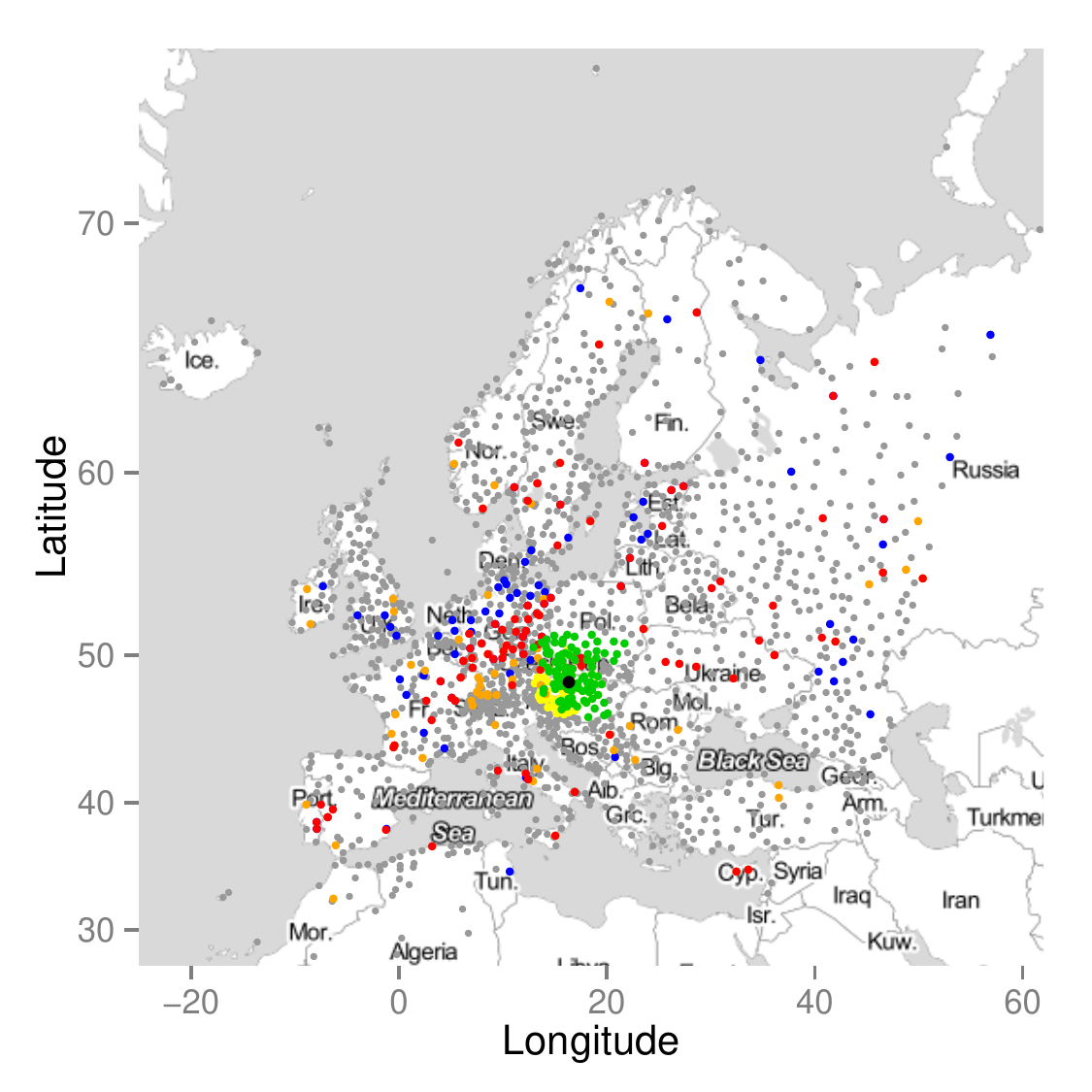}}\hfill \\
\subfigure{\includegraphics[width = 0.7\textwidth]{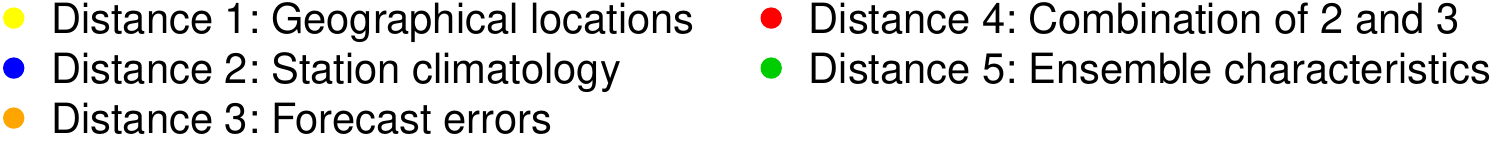}} 
\caption{Illustration of the 100 most similar stations measured by the five distance functions for two reference stations at Ouessant, France (a) and Vienna, Austria (b). The reference stations are indicated by black dots. Note that several points are part of the set of close stations in more than one distance measure. In this case, they are assigned the color of the last mentioned distance. See Figure \ref{fig:AppendxFig1} in the 
Appendix for individual plots.}
\label{fig:distances-illustration}
\end{figure}

Figure \ref{fig:distances-illustration} illustrates the five distance functions for two of the observation stations by displaying the 100 most similar stations in a specific color each. For both stations, a portion of the sets of most similar stations measured by two or more distance functions overlaps. See Figure \ref{fig:AppendxFig1} in the Appendix for individual plots for the five distance functions and the two stations. 

For the station at Ouessant (Figure \ref{fig:distances-illustration}a) which is located on the North-Western coast of France, it can be observed that the 100 most similar stations measured by the distance functions depending on the distribution of the observations and forecast errors (distances 2--4) are mostly located at coastal regions and islands in Northern Europe, in particular if these characteristics are combined (distance 4). By contrast, the most similar stations to the observation site at Vienna (Figure \ref{fig:distances-illustration}b) are distributed over continental central Europe, mostly located in France, Germany and Poland. 

As implied by the definition, the most similar stations measured by distance 1 (and due to the large overlap also by distance 5) are located in close geographical proximity around the two observation sites. Due to the differences in the density of the observation station network, the stations similar to the reference station at Ouessant are spread out over larger geographical distances compared to the respective stations similar to the one at Vienna. Therefore, data from stations with potentially significantly different climatological properties might be added to the training sets for parameter estimation.  

\subsubsection*{Clustering-based semi-local model}

Further, as an alternative to the distance-based approach we propose a novel semi-local approach based on cluster analysis. Here, the observation sites are grouped into clusters, and parameter estimation is performed  for each cluster individually using only ensemble forecasts and validating observations at stations within the given cluster. To determine the clusters of observation stations we apply $k$-means clustering \citep[see, e.g.,][]{hastie09} to various choices of feature sets which are based on climatological characteristics of the observation stations and the distribution of forecast errors, and are described in more detail below. 

In comparison to the distance-based method, the clustering-based semi-local approach is computationally much more efficient, as the parameter estimation is only performed for $k$ distinct training sets for each given day, whereas the distance-based approach requires individual estimation of the coefficients at each of the 1738 stations with partially overlapping training sets. Further, the similarities between the observation stations are obtained in a more efficient way as clustering is computationally less demanding compared to the computation of pair-wise distances between all observation stations (up to symmetry)\footnote{The number of distances that have to be computed in every training set is $\frac{1737\cdot1738}2 \approx 1.5\cdot 10^6$.}. In particular, clustering-based semi-local estimation is also computationally more efficient than local parameter estimation which arises as a special case with $k = 1738$ clusters of size 1 each. 

The above discussion does not account for the computational costs of the actual clustering. However, there exist efficient algorithms for $k$-means  clustering, e.g., the Hartigan-Wong algorithm \citep{hw79}, which converge rapidly for the data at hand. The costs of the actual clustering are thus negligible compared to the computational costs of the numerical parameter estimation.  

In contrast to the distance-based approach, this allows for iteratively determining the clusters anew in every training period without a significant increase in the overall computational costs. This adaptive approach will be pursued for all clustering-based semi-local models discussed below.

We denote the number of features used in the $k$-means clustering procedure by $N$ and consider the following feature sets.

{\em Feature set 1: Station climatology}. Let $\hat F_{i,n}$ denote the empirical CDF of the wind speed observations at station $i$ over the rolling training period consisting of the preceding $n$ forecast cases at this station. The feature set for station $i$ is given by the set of equidistant quantiles of $\hat F_{i,n}$ at levels $\frac{1}{N+1},\frac{2}{N+1},\dots,\frac{N}{N+1}$.

{\em Feature set 2: Forecast errors}. Denote the empirical CDF \eqref{eq:ECDFfcerrors} of forecast errors $e_{i,t}$ by $\hat G_{i,n}^e(z)$. With a slight abuse of the above notation, the set $T$ in the expression $t\in T$ denotes the preceding $n$ dates as the clusters are iteratively determined anew in every rolling training period. The feature set for station $i$ is then given by the set of equidistant quantiles of  $\hat G_{i,n}^e$ at levels $\frac{1}{N+1},\frac{2}{N+1},\dots,\frac{N}{N+1}$.

{\em Feature set 3: Combination of feature sets 1 and 2}. To define a feature set that depends on both the station climatology and the distribution of forecast errors, we combine equidistant quantiles of  $\hat F_{i,n}$ at levels $\frac{1}{N_1+1},\dots,\frac{N_1}{N_1+1}$ and equidistant quantiles of $\hat G_{i,n}^e$ at levels  $\frac{1}{N_2+1},\dots,\frac{N_2}{N_2+1}$ into one single set of size $N = N_1 + N_2$, where $N_1$ and $N_2$ are defined as follows. If $N$ is an even number, let $N_1 = N_2 = \frac{N}{2}$, otherwise let $N_1 = \lceil \frac{N}{2} \rceil$ and $N_2 = N - N_1$. 

Alternative choices of feature sets where the geographical location of the observation stations is included in the definition have also been investigated, but result in a reduction of the predictive performance and are thus omitted in the following discussion. 

\begin{figure}[t]
\centering
\subfigure[Climatology]{\includegraphics[width=0.33\textwidth]{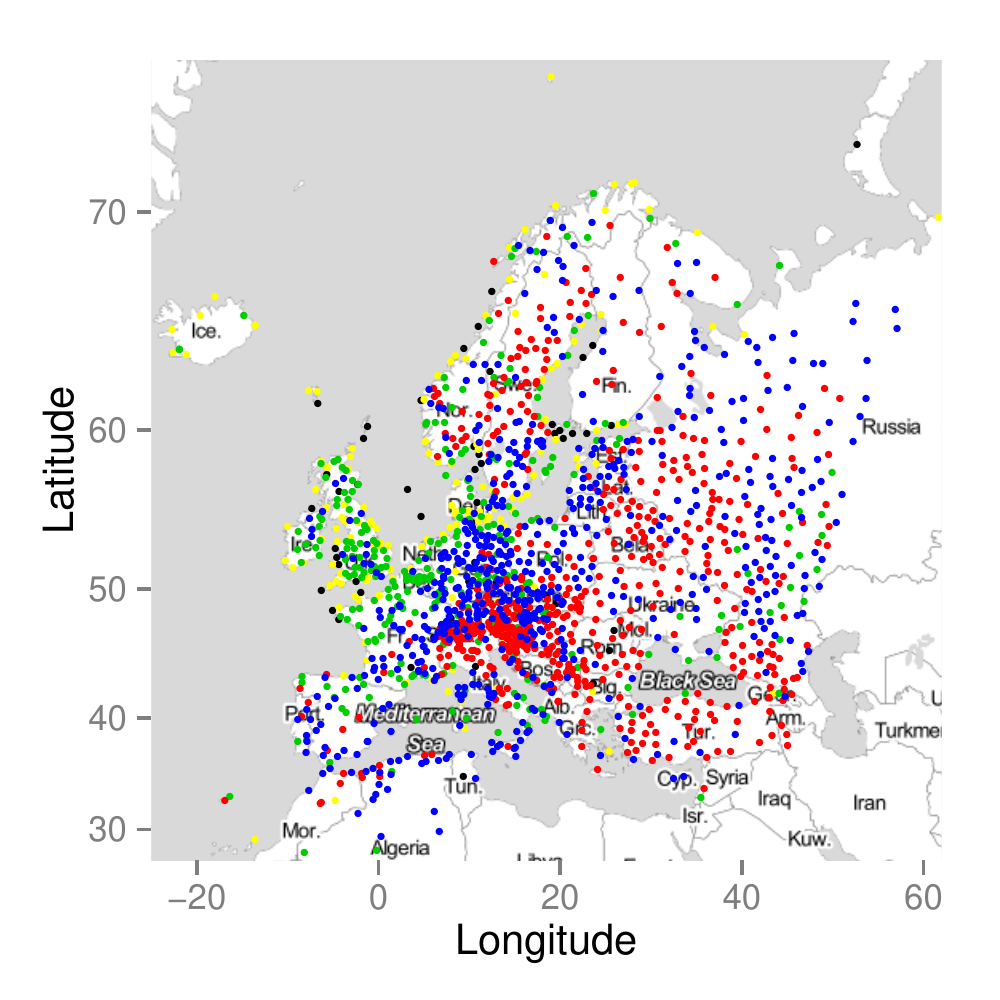}}\hfill  
\subfigure[Forecast errors]{\includegraphics[width=0.33\textwidth]{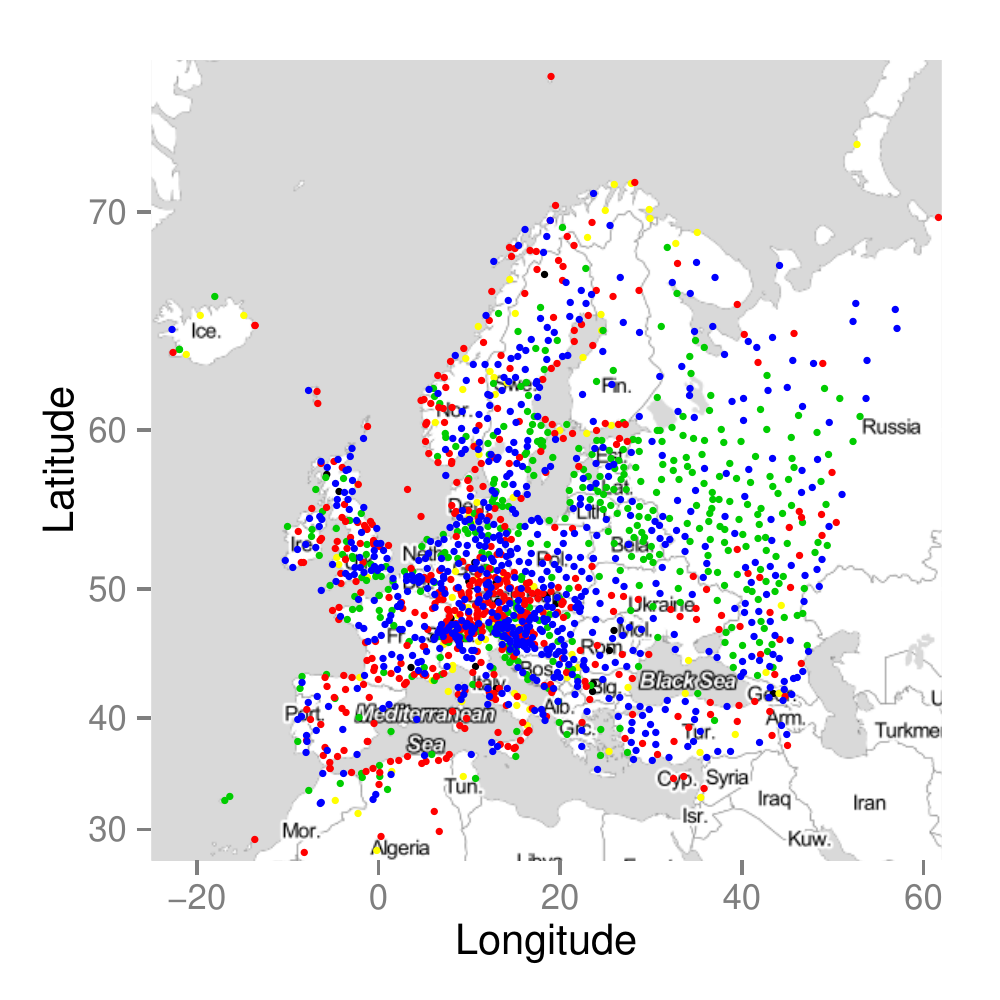}}\hfill  
\subfigure[Climatology + forecast errors]{\includegraphics[width=0.33\textwidth]{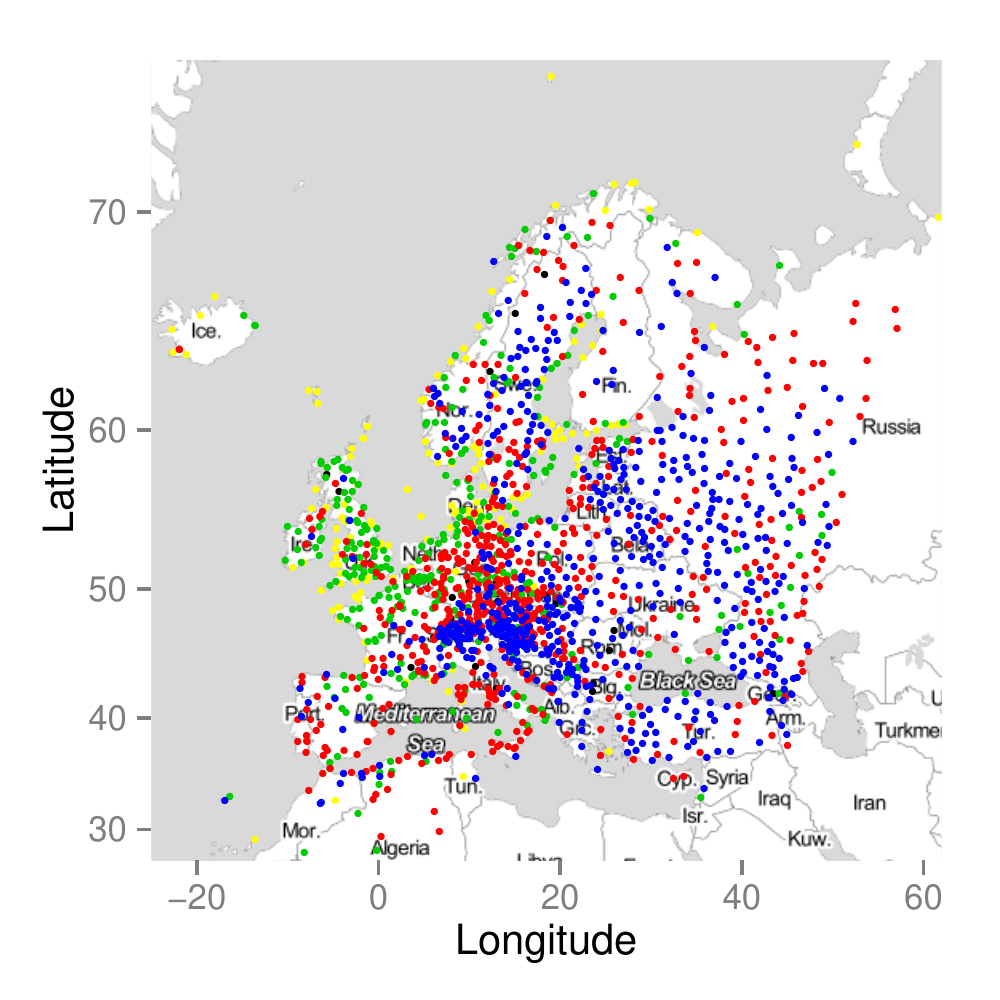}}\hfill 
\caption{Illustration of cluster memberships of the observation stations based on feature sets 1 (a), 2 (b) and 3 (c) obtained with a fixed number of 5 clusters and 24 features. Colors are assigned to the clusters by size (in descending order: blue, red, green, yellow, black).}
\label{fig:clustering-illustration}
\end{figure}

Figure \ref{fig:clustering-illustration} illustrates the obtained clusters of observation stations for the different feature sets with a fixed number of $k = 5$ clusters. For the feature set defined in terms of the distribution of the observations (feature set 1, Figure \ref{fig:clustering-illustration}a), one can observe two larger clusters distributed over central Europe, where one cluster mainly contains stations in Germany and France, while the other one contains most of the stations in the Alps and continental Eastern Europe. The remaining clusters are predominantly centered around the United Kingdom and coastal regions of France and Northern Europe. If the clusters are determined based on forecast errors (feature set 2, Figure \ref{fig:clustering-illustration}b), the stations are mainly grouped into three almost equally large clusters, where the most notable difference compared to the fist feature set is the predominant presence of the third cluster in North-Eastern Europe. Further, the stations in the 
United Kingdom and coastal regions of Europe now mostly belong to the two biggest clusters rather than forming separate sets. Clustering based on a combination of the distribution of the observations and forecast errors (feature set 3, Figure \ref{fig:clustering-illustration}c) results in a pattern of cluster memberships in between the other two choices. In particular, the alpine regions, continental Europe and the coastal regions and the United Kingdom show the most clear-cut separation compared to the other feature sets.

\section{Results}
  \label{sec:sec4}
  
\subsection{Model formulations}\label{subs:subs4.1}

As discussed in Section \ref{subs:subs3.1}, the link functions connecting the parameters of the predictive distribution of the EMOS models and the ensemble forecasts depend on the stochastic properties of the ensemble. The GLAMEPS ensemble consists of four subensembles which differ in the choice of numerical model and parametrization scheme. Each subensemble contains a control and $6+6$ (non-lagged and lagged) perturbed members. This induces a natural grouping into twelve groups: 
\begin{align*}
 f_{AI,1}, \dots, f_{AI,6} & \qquad \text{ALARO model with ISBA parameterization scheme} \\
 f_{AS,1}, \dots, f_{AS,6} & \qquad \text{ALARO model with SURFEX parameterization scheme} \\
 f_{HK,1},\dots, f_{HK,6} & \qquad \text{HIRLAM model with Kain-Fritsch parameterization scheme} \\
 f_{HS,1}, \dots, f_{HS,6} & \qquad \text{HIRLAM model with STRACO parameterization scheme} \\
 f_{\bullet L,1}, \dots, f_{\bullet L,6} & \qquad \text{lagged versions of above groups, 4 individual groups of size 6,} \\
 & \qquad\quad \text{where } \bullet \in \{ AI, AS, HK, HS\} \\ 
 f_{AI,c},f_{AS,c}, f_{HK,c}, f_{HS,c} & \qquad  \text{control forecasts, 4 individual groups of size 1.} 
\end{align*}

The members within each individual group are exchangeable and should share a common set of EMOS coefficients, resulting in a predictive TN distribution with location
\begin{align}
   \label{eq:eq4.1}
  a_0\!&+\!a_{AI,c}f_{AI,c}\!+\!\sum_{\ell_1=1}^{6}\!\big (a_{AI}f_{AI,\ell_1}
   \!+\! a_{AIL}f_{AIL,\ell_1}\big)
   \!+\!a_{AS,c}f_{AS,c}\!+\!\sum_{\ell_2=1}^{6}\!\big(a_{AS}f_{AS,\ell_2}
   \!+\!a_{ASL}f_{ASL,\ell_2}\big) \\
   &+\!a_{HK,c}f_{HK,c}\!+\!\sum_{\ell_3=1}^{6}\!\big (a_{HK}f_{HK,\ell_3}
   \!+\! a_{HKL}f_{HKL,\ell_3}\big)
   \!+\!a_{HS,c}f_{HS,c}\!+\!\sum_{\ell_4=1}^{6}\!\big(a_{HS}f_{HS,\ell_4}
   \!+\!a_{HSL}f_{HSL,\ell_4}\big) \nonumber
\end{align}
and scale $b_0+b_1S^2$, which is a special case of model \eqref{eq:eq3.3}. This model has a total number of $15$ parameters to be estimated and will be referred to as {\em full model}. 

A natural simplification is to assign the same parameter values to the lagged and non-lagged exchangeable ensemble members of a subensemble, which results in a reduced model with location 
\begin{align}
   \label{eq:eq4.2}
  a_0\!&+\!a_{AI,c}f_{AI,c}\!+\!\sum_{\ell_1=1}^{6}\!a_{AI}\big (f_{AI,\ell_1}
   \!+\! f_{AIL,\ell_1}\big)
   \!+\!a_{AS,c}f_{AS,c}\!+\!\sum_{\ell_2=1}^{6}\!a_{AS}\big(f_{AS,\ell_2}
   \!+\!f_{ASL,\ell_2}\big) \\
   &+\!a_{HK,c}f_{HK,c}\!+\!\sum_{\ell_3=1}^{6}\!a_{HK}\big (f_{HK,\ell_3}
   \!+\! f_{HKL,\ell_3}\big)
   \!+\!a_{HS,c}f_{HS,c}\!+\!\sum_{\ell_4=1}^{6}\!a_{HS}\big(f_{HS,\ell_4}
   \!+\!f_{HSL,\ell_4}\big) \nonumber
\end{align}
and $11$ parameters to be estimated. This model will be referred to as {\em lag-ignoring model}. 

Finally, we also investigate the fully exchangeable situation where the existence of the aforementioned groups is ignored, and all ensemble members are assumed to form a single exchangeable group. In this case the  predictive distribution is given by
\begin{equation}
   \label{eq:eq4.3}
{\mathcal N}_0\big(a_0+a_1\overline f,b_0+b_1S^2\big), 
\end{equation}
where again, $\overline f$ denotes the ensemble mean, and we refer to this model as {\em simplified model}.  
 
 \subsection{Selection of tuning parameters for semi-local parameter estimation methods}\label{sec:4-tuningpar}
 
 Both semi-local parameter estimation techniques require the choice of various tuning parameters given by the length of the rolling training period, the number of similar stations to be taken into account, the number of features and the number of clusters. We now discuss the effect of these tuning parameters on the predictive performance of the forecast models. To that end, the full, lag-ignoring and simplified model were estimated using the distance-based and clustering-based semi-local parameter estimation techniques described in Section \ref{subs:subs3.3}. Conclusions are drawn based on the mean CRPS over the evaluation period. For comparison, note that the average CRPS values of the GLAMEPS ensemble and the best regional TN model with a training period of 80 days are 1.058 and 0.955, respectively. 
 
 Due to numerical stability issues in the parameter estimation, a comparison to local TN models is impossible, an  estimate of the average CRPS of the locally estimated simplified TN model with a training period of 80 days can be obtained if the problematic parameter estimates (around 0.1\% of the total number of forecast cases) are replaced by corresponding estimates from preceding forecast cases. This estimate of the average CRPS of the local simplified model with such subsequent modifications equals 0.790 (see Section \ref{subs:subs4.3}).  
 
 \subsubsection*{Distance-based approach}
 
 \begin{figure}[t]
\centering 
\includegraphics[width=0.99\textwidth]{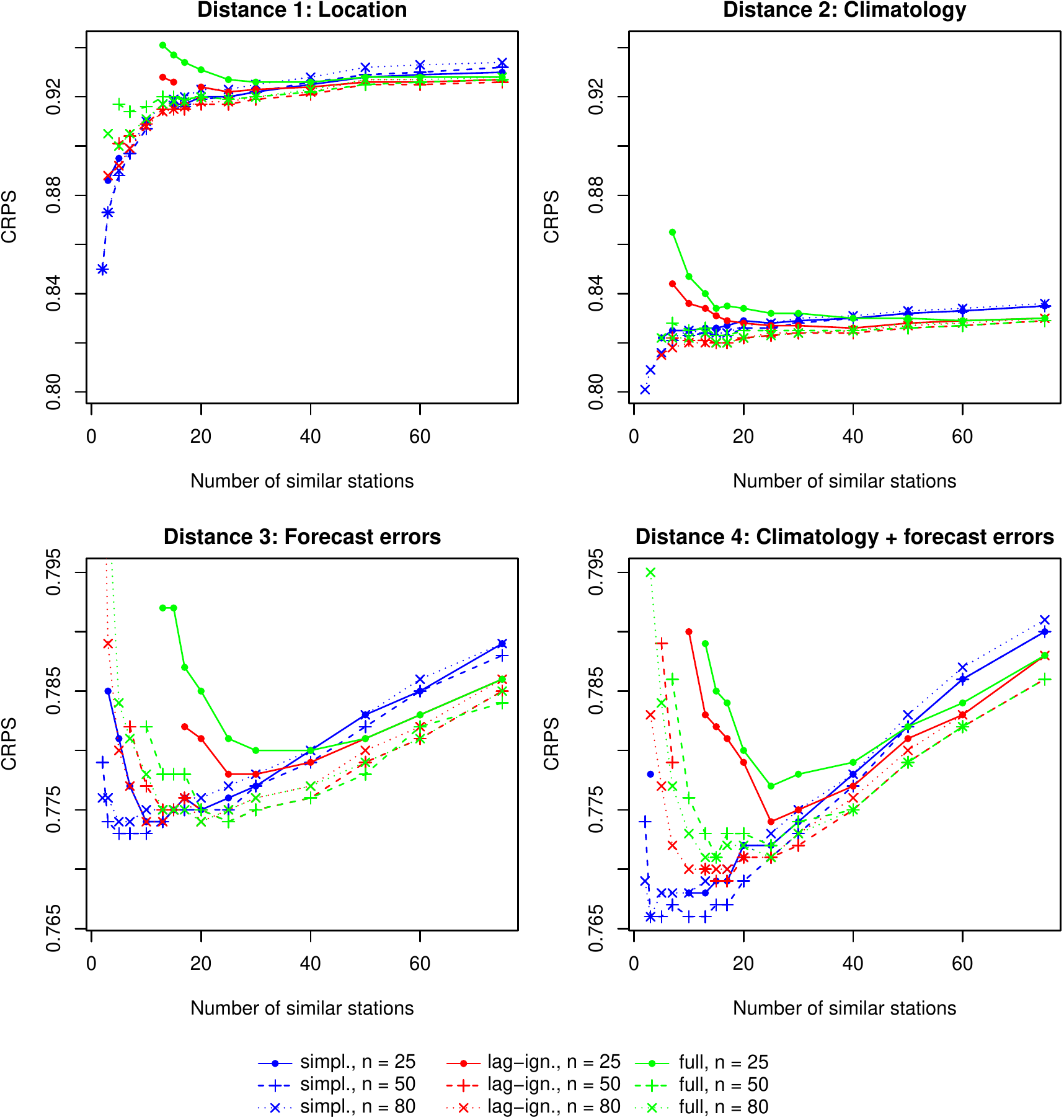}.
\caption{Effect of the number of similar stations $L$ on the predictive performance of the distance-based semi-local models for three choices of training period lengths $n$ (in days). Missing line segments indicate unsuccessful parameter estimation for these choices of tuning parameters.}
\label{fig:tuningpar-L-distbased}
\end{figure}
 
In the distance-based semi-local approach to parameter estimation, the size of the training set for a given station $i$ is increased by including corresponding training data from the $L$ most similar stations, i.e., the $L$ stations with the smallest distances $d(i,j),\,j\in\{1,\dots,1738\}$. Note that for the distance functions defined in Section \ref{subs:subs3.3}, $d(i,i) = 0$, a value of, e.g, $L = 5$ thus means that the training set for station $i$ consists of data from this station, and of data from the 4 stations with the smallest distances to station $i$. Figure \ref{fig:tuningpar-L-distbased} illustrates the effect of the number of close stations on the predictive performance measured as mean CRPS of the three proposed models for selected lengths of the training period. Due to the large overlap of close stations determined by distance functions 1 and 5 (see, e.g., Figure \ref{fig:distances-illustration}) we omit the corresponding plots for distance 5 which closely resemble the plots for distance 1 
and remark that similar conclusions apply, in particular for small values of $L$. Note the varying scales of the plots in the first and second row of Figure \ref{fig:tuningpar-L-distbased} caused by the different predictive performances of the respective models.

For distance 1 which is based on geographical locations, the predictive performance generally decreases with the number of similar stations added to the training sets, except for the more complex lag-ignoring and full models and shorter training periods, where the best CRPS values are attained for values around $L = 20$. Clearly, the inclusion of similar stations then allows for unproblematic parameter estimation, but as few stations as possible should be chosen in order to achieve results as close as possible to the desirable (however, even for long training periods impossible) local parameter estimation. Similar conclusions apply for the climatology-based distance  2, however, the predictive performance of these models is notably better. 

A different pattern emerges for distance 3 which is based on the distribution of forecast errors. Particularly for the more complex lag-ignoring and full model, the best predictive performances are achieved with choices of $L$ between 10 and 30, depending on the length of the training periods, whereas smaller values of $L$ result in worse predictions. Note that with these choices of $L$, the predictive performance of the semi-local models is better than the estimate of the predictive performance for the (simplified) local model. For distance  4, a combination of distance functions 2 and 3, similar conclusions apply with optimal values of $L$ between 10 and 25. Semi-local models based on this similarity measure show the best predictive performance and are also able to outperform the simplified local TN model for a wide range of tuning parameter choices.

The effect of the length of the rolling training periods consisting of the preceding $n$ days can also be seen from Figure \ref{fig:tuningpar-L-distbased} where each individual plot contains three different choices of $n$. Together with further investigations of plots of the average CRPS against the employed training period lengths (not shown), one can observe that $n$ only has a small effect on the predictive performance of the models.

For all considered distance functions, the predictive performance increases with longer training periods, in particular for the more complex models and smaller values of $L$. This is to be expected from the smaller size of the training sets as parameter estimation becomes problematic for shorter training periods and few additional forecast cases from similar stations taken into account. 

\clearpage

The simplified models show a slight decrease in predictive performance for training periods longer than 40--50 days, however, the differences are negligible compared to those between  models based on varying choices of distance functions or varying numbers of similar stations taken into account. The overall best predictive performances across the three considered model formulations are achieved with training period lengths of 80 days. 

\subsubsection*{Clustering-based approach}
 
\begin{figure}
 \centering
 \includegraphics[width=\textwidth]{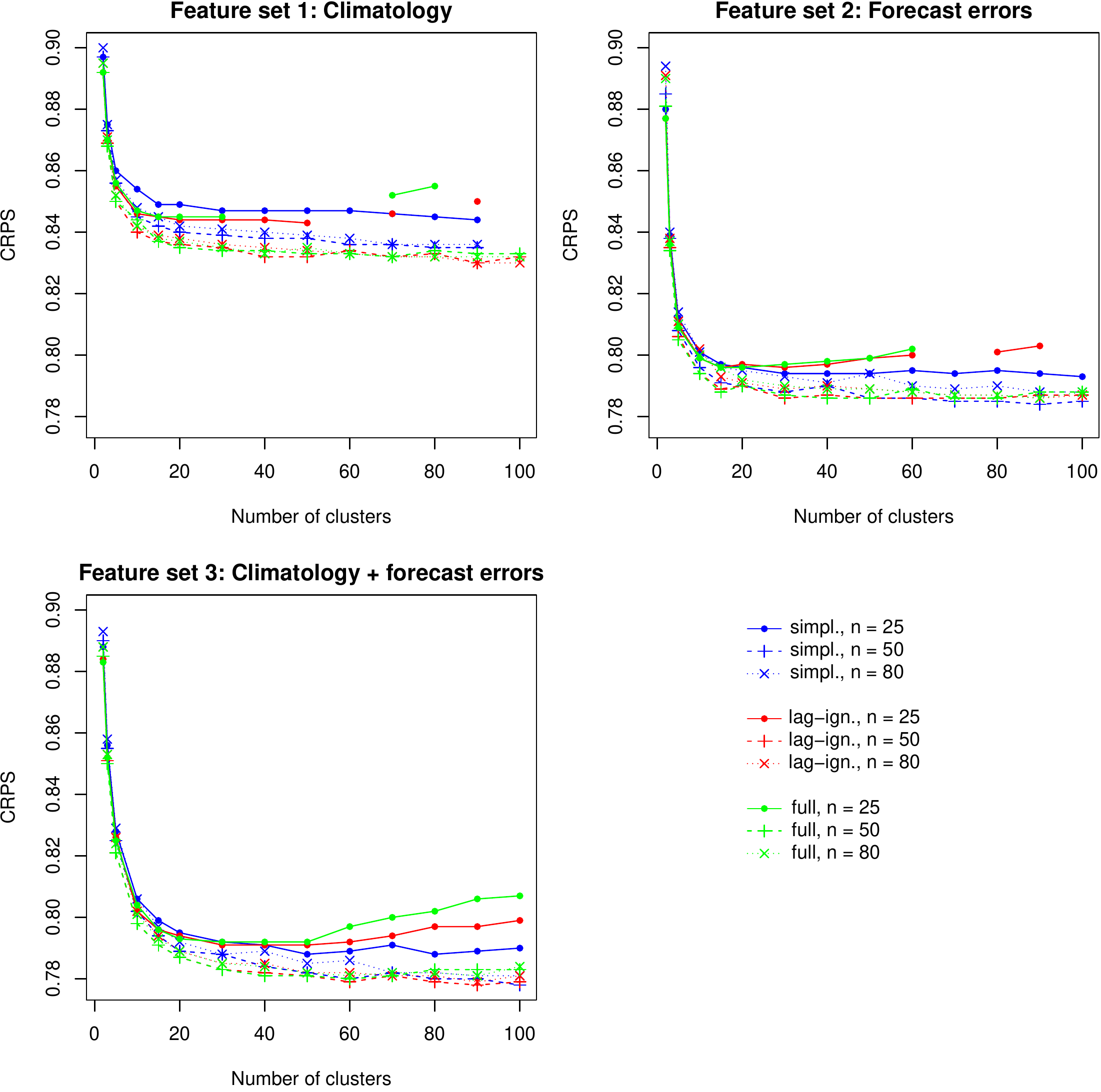}
 \caption{Effect of the number of clusters $k$ on the predictive performance of clustering-based semi-local models for three choices of training period lengths $n$ (in days). All models are estimated with feature sets of size $N = 24$. Missing line segments indicate unsuccessful parameter estimation for these choices of tuning parameters.}
\label{fig:tuningpar-k-clustering}
\end{figure}
 
 In the clustering-based semi-local approach $k$-means clustering based on the different feature sets (discussed in Section \ref{subs:subs3.3}) is employed to group the observation stations into clusters. The lower computational costs of this approach allow for iterative computation of the clusters in every training period, whereas the similarities between stations used in the distance-based semi-local approach are computed over a fixed period of data from October 2013 to February 2014 preceding the verification period. This adaptive application of $k$-means clustering leads of improvements in mean CRPS of around 1-5\% compared to the use of a fixed set of clusters determined over the first period of available data. 
 
 Figure \ref{fig:tuningpar-k-clustering} illustrates the effect of the number of clusters $k$ on the predictive performance of the clustering-based semi-local models. Choosing $k = 1$ obviously corresponds to regional parameter estimation. For all three feature sets considered here, the predictive performance increases for larger values of $k$ up to around 100 clusters except for shorter training periods. Clearly, a larger number of clusters allows for a more refined grouping into sets of observation stations with similar characteristics. The predictive performance decreases for all considered models and training period lengths if much more than $k=100$ clusters are used. This behavior is to be expected as the clusters become smaller and parameter estimation becomes numerically unstable, particularly for the lag-ignoring and full models. Note that depending on training period length and feature set, only small improvements can be observed for $k$ exceeding values of around 40 to 70 clusters.
 
 As observed for the distance-based models, the clustering-based semi-local models defined in terms of the distribution of forecast errors and the station climatology (feature sets 2 and 3) are able to outperform the local model over a wide range of tuning parameter choices except for short training periods. The worse predictive performance for shorter training periods is to be expected as the smaller amount of forecasts cases used to determine the clusters might result in a less accurate partitioning of the observation stations. Compared to the distance-based approach it can be observed that for some numbers of clusters, training period lengths below 80 days are optimal, in particular for the lag-ignoring and full model. However, in comparison to the effect of different choices of feature sets the effect of the length of the training period is negligible. 
 
 \begin{figure}
 \centering
 \includegraphics[width=\textwidth]{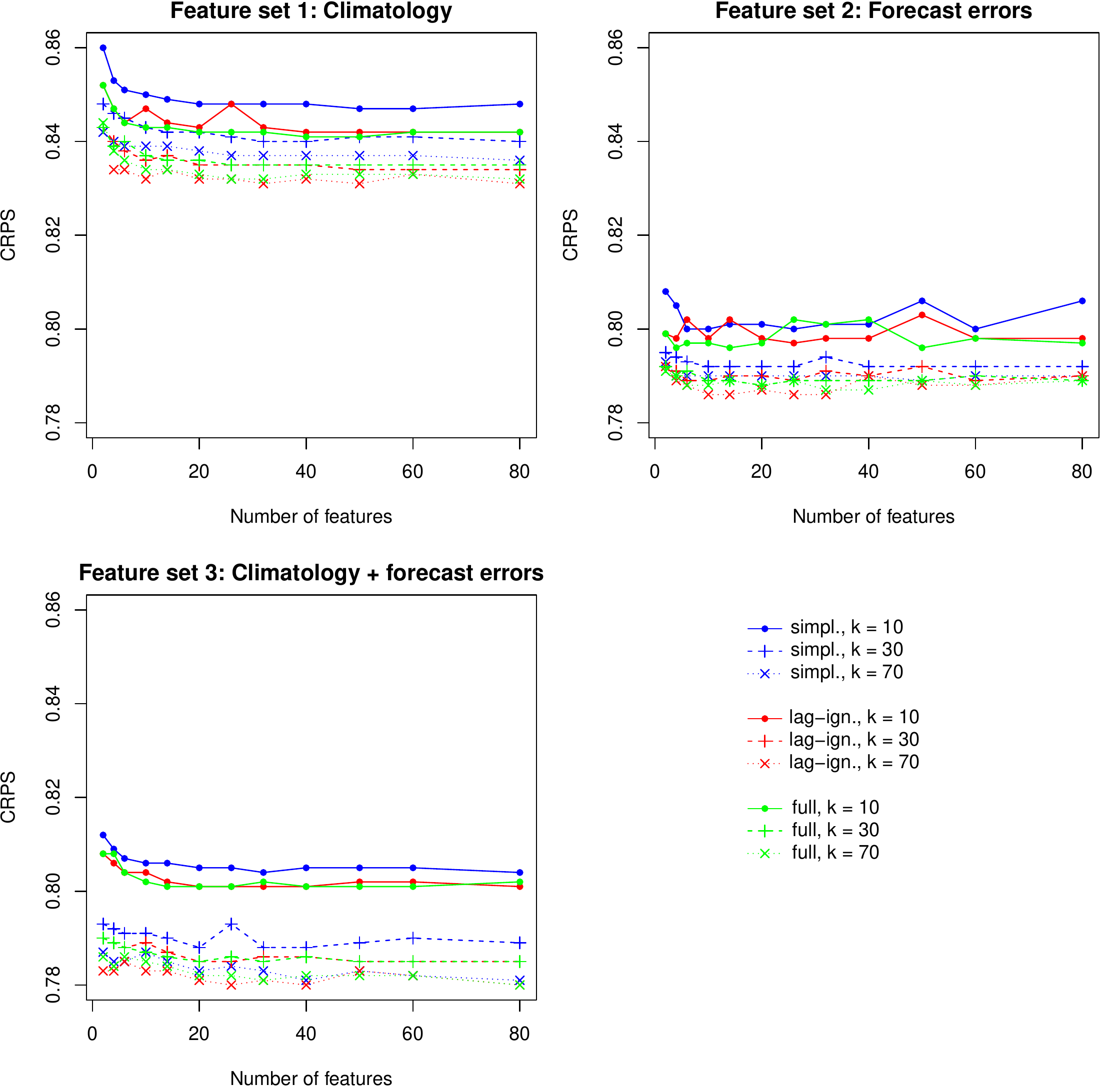}
 \caption{Effect of the size of the feature set $N$ on the predictive performance of clustering-based semi-local models for three choices of numbers of clusters $k$. All models are estimated over a training period of 80 days. Missing line segments indicate unsuccessful parameter estimation for these choices of tuning parameters.}
\label{fig:tuningpar-N-clustering}
\end{figure}
  
 Thus far, all clustering-based semi-local models shown in Figure \ref{fig:tuningpar-k-clustering} were estimated for a fixed feature set size of $N = 24$. To illustrate the effect of $N$ on the predictive performance, Figure \ref{fig:tuningpar-N-clustering} shows the average CRPS of the clustering-based models as functions of the number of features $N$ considered in $k$-means clustering for three choices of $k$. Given that sufficiently many features (around 5-10 depending on the other tuning parameters) are used, the feature set size has only a small effect on the predictive performance compared to different choices of $k$ or $n$. Reasons for this behavior clearly include the aforementioned robustness of the obtained cluster memberships with regards to $N$. The best results across all considered tuning parameter combinations are generally obtained for feature set sizes between 20 and 40 thus justifying our previous choice of $N = 24$.

 \subsection{Forecast performance}\label{subs:subs4.3}
 
 The predictive performance of the semi-local models is evaluated by computing the verification scores introduced in Section \ref{subs:subs3.2} over the verification period March 1 -- May 18, 2014. We use the local climatological forecasts given by the observations at the corresponding station during the rolling training periods, the raw GLAMEPS ensemble predictions, and probabilistic forecast by the regional TN model as benchmark models. While locally estimated models are desirable, the estimation of these models is highly problematic for the GLAMEPS data due to the issues discussed earlier. Even for the simplified model \eqref{eq:eq4.3} with a maximum training period length of 80 days, numerical issues occur in the local parameter estimation, e.g., some shape parameters are estimated to be 0. An estimate of the predictive performance of the local model can be obtained by replacing these problematic parameter estimates by the preceding ones. However, note that these subsequent adjustments are not necessary 
for the semi-local or regional models. Further, neither the lag-ignoring nor the full local TN model can be successfully estimated as the employed numerical optimization algorithms fail to converge or produce numerical errors.   
 
 In the interest of brevity, we limit our discussion to the simplified and the lag-ignoring models. It can be seen from Figures \ref{fig:tuningpar-L-distbased}--\ref{fig:tuningpar-N-clustering} that the full semi-local models generally result in slightly worse predictive performance compared to the lag-ignoring models, therefore the additional computational costs of taking into account the lagging in the subensembles are not justified. Note that different conclusions may apply for other ensemble prediction systems with lagged members.
 
 With regards to the tuning parameters for the semi-local approaches, we employ a fixed training period length of 80 days, and use a fixed number of  $N = 24$ features for $k$-means clustering to ensure comparability across the different models. For the individual distance-based and clustering-based semi-local models we then choose suitable values for the number of most similar stations $L$ and the number of clusters $k$ from Figures \ref{fig:tuningpar-L-distbased}--\ref{fig:tuningpar-N-clustering} (see Section \ref{sec:4-tuningpar} for a detailed discussion of the effect of these tuning parameters). While the chosen tuning parameter combinations might not be the overall optimal values for the individual models, the results hold for a wide range of tuning parameter choices as indicated by the sensitivity considerations in Section \ref{sec:4-tuningpar}.
 
  \begin{table}
  \centering
  \small
  \caption{Mean CRPS, MAE, coverage and width of 96.2\% prediction intervals of probabilistic 18h ahead forecasts of wind speed evaluated over the second period of data from March to May 2014. A training period length of 80 days is used for all models. For the clustering-based model estimation, a fixed number of $N = 24$ features is applied.} 
  \begin{tabular}{l@{\hskip -1cm}l@{\hskip 1.5cm}cccc}
    \toprule
		& & CRPS         & MAE          & Coverage & Width \\
      Forecast 	& & (m s$^{-1}$) & (m s$^{-1}$) & (\%)     & (m s$^{-1}$) \\
    \midrule
      Local climatology & & 1.127 & 1.580 & 96.6 & 7.96 \\
      GLAMEPS ensemble  & & 1.058 & 1.376 & 67.1 & 3.50 \\
    \midrule
      \multicolumn{5}{l}{\textit{Regional TN models}}  \\
    \midrule
    simpl.   & & 0.957 & 1.324 & 90.3 & 6.36 \\
    lag-ign. & & 0.955 & 1.320 & 90.3 & 6.33 \\
    \midrule
    \multicolumn{5}{l}{\textit{Local TN models (with subsequent modifications)}}  \\
    \midrule
    simpl.   & & 0.790 & 1.100 & 88.7 & 5.12 \\
    \midrule
    \multicolumn{5}{l}{\textit{Distance-based semi-local TN models}} & \\
    \midrule
    D1 simpl.  & $L = 3$  & 0.873 & 1.218 & 90.2 & 5.99 \\
    D1 lag-ign.& $L = 3$  & 0.887 & 1.236 & 89.2 & 5.71 \\
    D2 simpl.  & $L = 5$  & 0.816 & 1.136 & 90.0 & 5.61 \\
    D2 lag-ign & $L = 5$  & 0.815 & 1.136 & 89.6 & 5.42 \\
    D3 simpl.  & $L = 5$  & 0.774 & 1.083 & 90.3 & 5.25 \\
    D3 lag-ign.& $L = 10$ & 0.774 & 1.083 & 90.2 & 5.21 \\
    D4 simpl.  & $L = 3$  & 0.766 & 1.069 & 89.9 & 5.16 \\
    D4 lag-ign.& $L = 10$ & 0.770 & 1.075 & 90.0 & 5.18 \\
    D5 simpl.  & $L = 3$  & 0.874 & 1.220 & 90.2 & 5.95 \\
    D5 lag-ign.& $L = 5$  & 0.895 & 1.248 & 89.8 & 5.91 \\
    \midrule
    \multicolumn{5}{l}{\textit{Clustering-based semi-local TN models}} & \\
    \midrule
    C1 simpl.   & $k = 70$ & 0.836 & 1.162 & 89.8 & 5.68 \\
    C1 lag-ign. & $k = 70$ & 0.832 & 1.156 & 89.6 & 5.55 \\
    C2 simpl.   & $k = 70$ & 0.789 & 1.103 & 89.9 & 5.25 \\
    C2 lag-ign. & $k = 70$ & 0.787 & 1.099 & 89.8 & 5.22 \\
    C3 simpl.   & $k = 70$ & 0.782 & 1.091 & 89.7 & 5.19 \\
    C3 lag-ign. & $k = 70$ & 0.781 & 1.090 & 89.7 & 5.17 \\
    \bottomrule
    \end{tabular}
    \label{table:perf-measures}
  \end{table}
    
 Table \ref{table:perf-measures} shows the average CRPS, MAE of median values, and coverage and average width of 96.2\% prediction intervals for the considered models. The raw GLAMEPS ensemble predictions outperform the climatological forecasts and provide sharp prediction intervals, however, at the cost of being uncalibrated. Regional TN models are able to improve the calibration of the ensemble, and result in around 10\% better mean CRPS values, however, the semi-local approaches significantly outperform the regional approaches for all considered models and tuning parameter choices, see also Figures \ref{fig:tuningpar-L-distbased} and \ref{fig:tuningpar-k-clustering}. 
  
 Among the distance-based semi-local models, the best predictive performances are obtained by distance functions 3 and 4 which utilize the distribution of forecast errors and combinations with the station climatology to determine similarities between stations. Note that these semi-local models are also able to outperform the local TN model for a wide range of tuning parameter choices without requiring subsequent corrections and while further allowing for a successful estimation of the more complex lag-ignoring and full semi-local models. The semi-local models based on distance functions 1 and 5 exhibit similar predictive performances which are slightly worse compared to the other distances, but are still able to outperform the regional model. The similarity is clearly caused by the large overlap of selected similar stations, see Figure \ref{fig:distances-illustration}. Except for distance 2, the simplified model \eqref{eq:eq4.3} performs slightly better than the lag-ignoring model \eqref{eq:eq4.2}, however, 
the differences are negligible compared to the differences between the different model estimation approaches.
 
 We obtain similar results for the clustering-based semi local models which perform slightly worse compared to the corresponding distance-based models, however, still significantly outperform the regional models and the local model if the clusters are determined on the basis of forecast errors and station climatology. Here, the lag-ignoring models show better predictive performances compared to the simplified models, but again, the differences are small compared to the influence of the choice of feature sets.
 
  \begin{figure}[h]
  \centering
  \includegraphics[width=\textwidth]{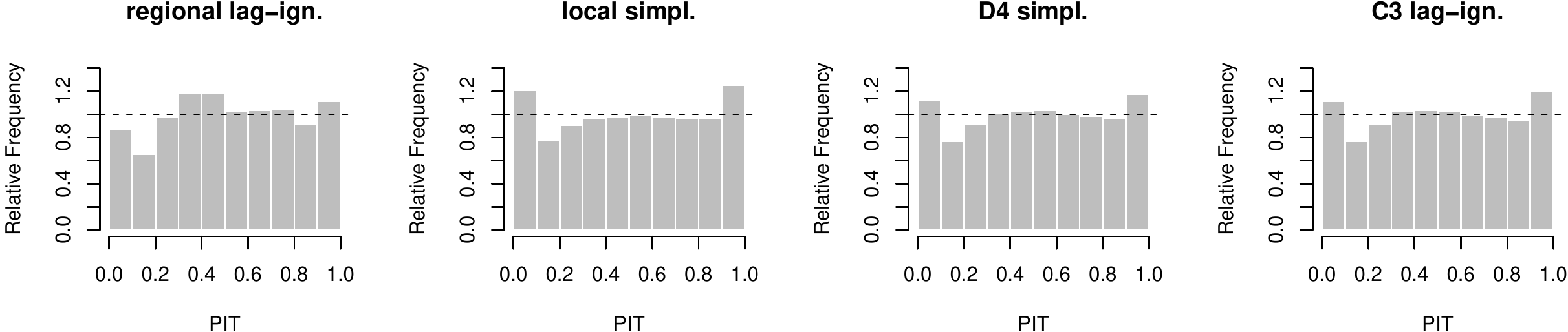} 
  \caption{PIT histograms of the EMOS postprocessed forecasts. All models are estimated with a rolling training period of 80 days. The displayed semi-local models are those with the best mean CRPS, see Table \ref{table:perf-measures} for the corresponding tuning parameter choices.}
  \label{fig:PITs}
  \end{figure} 

 Figure \ref{fig:PITs} shows PIT histograms of the lag-ignoring regional, the simplified local, and the distance-based and clustering-based semi-local models with the best average CRPS values (see Table \ref{table:perf-measures}). Compared to the verification rank histogram of the raw GLAMEPS ensemble forecasts (see Figure \ref{fig:fig1}b), all postprocessing models exhibit significantly improved calibration with PIT histograms showing much smaller deviations from the desired uniform distribution. The hump-shaped PIT histogram of the regional TN model indicates a slight under-prediction of lower wind speed values. The local and semi-local models are able to correct for this deficiency and show slightly better calibration, in particular for the semi-local models. Most of the models in Table  \ref{table:perf-measures} show similarly shaped PIT histograms. Alternative distributional choices such as log-normal or generalized extreme value distributions might lead to further improvement in calibration, see e.g. 
 \citet{bl, bl15mix}.
 
 To conclude, we note that the overall best predictive performance is achieved by distance-based semi-local models utilizing both the distribution of observations as well as the distribution of forecast errors at the observation stations, closely followed by clustering-based models with feature sets defined in a similar way. These models show better predictive performances than the local model, and can be estimated without any numerical issues. Figures \ref{fig:tuningpar-L-distbased} and \ref{fig:tuningpar-k-clustering} indicate that these conclusions hold for a wide range of tuning parameter choices. With regards to the two semi-local approaches, the respective distance-based models generally show slightly better predictive performance, however, the estimation of the clustering-based models is computationally much more efficient and allows for an iterative application of the clustering algorithm in each training period.

\section{Discussion}
  \label{sec:sec5}
  
 We have proposed two semi-local approaches to parameter estimation for ensemble postprocessing where the training data for a given observation station are augmented with data from stations with similar characteristics. The distance-based approach roughly follows the ideas of \citet{hhw} and uses distance functions to determine the similarities between observations stations, whereas the novel clustering-based approach employs $k$-means clustering to obtain groups of similar stations. Various choices of distance functions, feature sets and tuning parameters have been tested. 
 
 The best results are obtained for semi-local models where the similarities between stations are determined based on combinations of the climatological distribution of observations as well as the distribution of forecast errors at the given stations. While all semi-local models show significantly better predictive performance than the regional models, these best models are also able to outperform the locally estimated model. The semi-local parameter estimation methods further allow for estimating more complex models without numerical issues, whereas local estimation is only possible for simplified model formulations with a reduced number of parameters and still requires subsequent modifications. 
  
 The semi-local models thus offer several advantages over the standard approaches to parameter estimation and are straightforward to implement. The clustering-based semi-local model estimation is further computationally much more efficient than local model estimation which arises as a special case with $k = 1738$ clusters of size 1 each. While distance-based semi-local models show slightly better predictive performance compared to the clustering-based models, the estimation requires substantially more computational resources. In particular, an adaptive computation of the similarities in every training period is not feasible for the distance-based models. 
 
 Compared to the work of \citet{hhw}, we propose several alternative distance functions and use the distance-based approach for observations at specific stations instead of gridded data. It would be interesting to apply the novel similarity measures as well as the clustering-based approach to grid-based forecast and analysis data and assess potential differences. In particular, distance functions incorporating the distribution of forecast errors (distances 3 and 4) result in significantly better predictive performance for the GLAMEPS data and might also offer improvements over the climatology-based distance function used by \citet{hhw} (similar to distance function 2) when applied to gridded data. 
 
 With regards to the results for the employed distance functions it might appear somewhat surprising that models based on similarities defined by characteristics of the ensemble (mean and variance) as measured by distance 5 do not result in improvements compared to simple location-based similarities (distance 1). However, this might be due to the fact that these characteristics of the ensemble are substantially influenced by the locations of the stations, and the training sets thus largely overlap with those of the location-based distance 1. These results might change for other ensemble prediction systems. Further, potential improvements might be obtained by including different summary statistics of the ensemble, e.g., by adding information about the within-group variances of the subensembles, or quantiles of the distribution of ensemble forecasts. 
 
 The group memberships of the observation stations in the clustering-based semi-local models are all determined by applying $k$-means clustering. Alternative clustering methods exist and might potentially lead to improvements \citep[for reviews and comparisons see, e.g.,][]{fr98, kr09}. We did not incorporate informations on the geographical locations of the stations or characteristics of the ensemble into the selected feature sets as initial tests indicated a worse predictive performance. For different ensemble prediction systems, these alternative choices of feature sets may lead to further improvements.
 
 In the interest of brevity, we limited our discussion to the standard truncated normal EMOS model proposed by \citet{tg}. An extension of the similarity-based semi-local parameter estimation approach to other postprocessing models might in particular be interesting for complex models where larger numbers of parameters have to be estimated and local parameter estimation might thus not be feasible \citep[for recent examples see, e.g.,][]{fst15, mtlg15, bl15mix}.
 
 \citet{jdma15} propose analog-based local EMOS models where the training set for a given station is chosen by selecting forecast cases with similar ensemble forecasts for that station. This analog-based approach thus utilizes information for a given station in an optimal way by selecting subsets of the local training sets, whereas our semi-local models combine informations from multiple observation stations based on similarities. While the analog-based modification of the local parameter estimation method shows good predictive performance in a case study on hub height wind speed, it requires sufficiently long training periods for locally selecting similar forecast cases. The implementation of this analog-based approach is thus infeasible for the GLAMEPS data, however, comparisons and combinations with the similarity-based semi-local approaches proposed here are of interest and might result in further improvement in predictive performance.

\bigskip
\noindent
{\bf Acknowledgments.} \ 
Sebastian Lerch gratefully acknowledges support by the Volkswagen Foundation through the program ``Mesoscale Weather Extremes -- Theory, Spatial Modelling and Prediction (WEX-MOP)'' and by Deutsche Forschungsgemeinschaft (DFG) through the Research Training Group ``RTG 1953 -- Statistical Modeling of Complex Systems and Processes''. 
S\'andor Baran was supported by the J\'anos Bolyai Research Scholarship of the Hungarian Academy of Sciences. The authors are indebted to Tilmann Gneiting and Michael Scheuerer for useful suggestions and remarks, and to Maurice Schmeits, Jan Barkmeijer and John Bj\o{}rnar Bremnes for providing the GLAMEPS data and assistance in data handling.

\newpage 

\begin{figure}[t]
\centering
\includegraphics[width = 0.5\textwidth]{distances-legend.pdf} \\
\includegraphics[height=0.87\textheight]{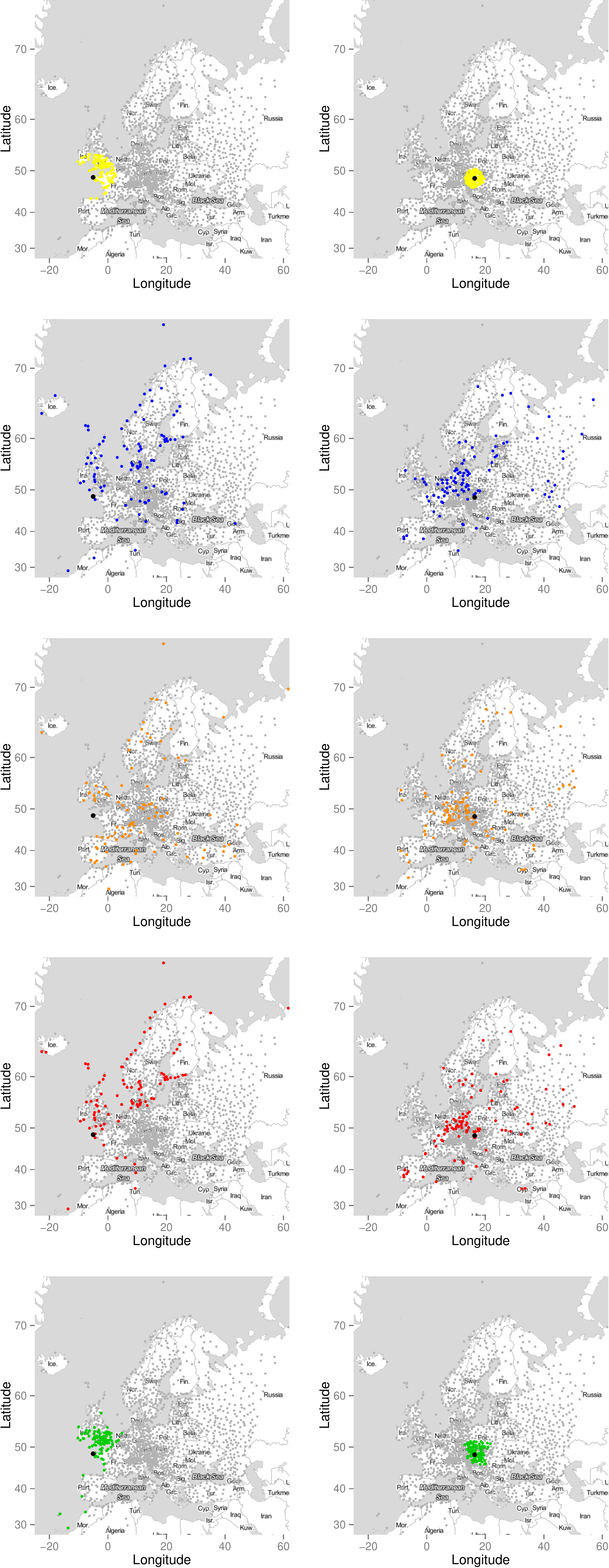} 
\caption{Appendix. Illustration of the 100 most similar stations measured by the five distance functions for two reference stations at Ouessant, France (left column) and Vienna, Austria (right column).}
\label{fig:AppendxFig1}
\end{figure}

\end{document}